\documentstyle[11pt,epsfig,aasms4]{article}
\begin{document}
\baselineskip 18.0pt
\def\oneskip{\vskip\baselineskip}
\def\xr#1{\parindent=0.0cm\hangindent=1cm\hangafter=1\indent#1\par}
\def\la{\raise.5ex\hbox{$<$}\kern-.8em\lower 1mm\hbox{$\sim$}}
\def\ma{\raise.5ex\hbox{$>$}\kern-.8em\lower 1mm\hbox{$\sim$}}
\def\ea{\it et al. \rm}
\def\am{$^{\prime}$\ }
\def\as{$^{\prime\prime}$\ }
\def\msol{M$_{\odot}$ }
\def\kms{$\rm km\, s^{-1}$ }
\def\cm3{$\rm cm^{-3}$}
\def\Ts{$\rm T_{*}$}
\def\F{$\rm F_{\nu}$}
\def\Vs{$\rm V_{s}$}
\def\n0{$\rm n_{0}$}
\def\B0{$\rm B_{0}$}
\def\ne{$\rm n_{e}$}
\def\Te{$\rm T_{e}$}
\def\Tgr{$\rm T_{gr}$}
\def\Tgas{$\rm T_{gas}$}
\def\Ec{$\rm E_{c}$}
\def\erg{$\rm erg\, cm^{-2}\, s^{-1}$}
\def\Hb{H$\beta$}
\def\upa{$\uparrow$}
\def\dop{$\downarrow$}

\bigskip

\bigskip

\bigskip

\bigskip

\centerline{\Large{\bf A Multi-Cloud Warm-Absorber Model for NGC 4051}}

\bigskip

\bigskip

\bigskip

\centerline{ $\rm M. \, Contini^1 \,\, and \,\,\, S. M. \, Viegas^2  $}

\bigskip

\bigskip

\bigskip

$^1$ School of Physics and Astronomy, Tel-Aviv University, Ramat-Aviv, Tel-Aviv,
69978, Israel

$^2$ Istituto Astron\^{o}mico e Geof\'{i}sico, USP, Av. Miguel Stefano, 
4200,04301-904
S\~{a}o Paulo, Brazil

\bigskip

\bigskip

\bigskip

\bigskip

\bigskip

\bigskip

\bigskip

Running title : Warm-Absorber Model for NGC 4051

\bigskip

\bigskip

\bigskip

\bigskip

\bigskip

\bigskip

subject headings : galaxies : nuclei - galaxies : Seyfert - shock waves -

galaxies : individual : NGC 4051 - X-ray : galaxies

\newpage

\section*{Abstract}

A multi-cloud model is presented which  explains  the soft X-ray excess
in  NGC 4051 and, consistently,  the
optical line spectrum and the SED of the continuum.
The clouds are heated and ionized by the photoionizing
flux from the active center and by shocks.
Diffuse radiation, partly absorbed throughout the clouds,
nicely fits the bump in the soft X-ray domain,
while bremsstrahlung radiation from the gaseous clouds
contribute to the fit of the continuum SED.
Debris of high density fragmented clouds are necessary to explain
the absorption oxygen throats observed at 0.87 and 0.74 keV.
The debris are heated by shocks of about 200-300 \kms.
Low velocity ($\leq$ 100 \kms) - density (100 \cm3) clouds contribute to the 
line and
continuum spectra, as well as high velocity (1000 \kms) - density (8000 \cm3) 
clouds
which are revealed by the FWHM of the line profiles.
The SED in the IR is explained by reradiation of
dust, however,  the dust-to-gas ratio is not particularly high ($\leq 
3\times 10^{-15}$).
Radio emission is well fitted by synchrotron radiation created
at the shock front by Fermi mechanism.

\newpage

\section{Introduction}

NGC 4051 is a SAB galaxy at z=0.0023. It is classified as a Seyfert 1 galaxy, 
and 
is characterized by unusually narrow permitted lines, only
moderately wider than the forbidden lines (Osterbrock 1977).

Simultaneous observations by ROSAT-IUE and GINGA
 of a sample of 8 Seyfert 1 galaxies (Walter et al 1994), 
including NGC 4051, show that the
 UV to X-ray spectral energy distribution (SED)  can be 
decomposed into two major distinct components:
 a nonthermal hard X-ray continuum and a broad emission excess (bump)
spanning from UV to soft X-rays.
All models (power-law, thin disk, bremsstrahlung, black body) 
are able to reproduce the soft X-ray 
spectra in the Walter et al. sample, except the power-law model for NGC 4051.
The evidence that the power-law model is not a good  representation
 of the X-ray spectrum of NGC 4051, 
when the absorbing column density is fixed to the galactic value, contrasts with 
the long established situation that
  the power-law is the best $simple$ fit to X-spectra of active galactic nuclei (AGN). 
This indicates that the spectrum is
more complicated, and, for example, could be  affected by intrinsic absorption.
 The addition of a soft component to the X-ray spectrum, 
which accounts for an excess in the 0.1-2 keV band, 
is also raised by Fiore et al. (1992). From the study of
GINGA data they found a spectral variability
consistent with a constant underlying power-law slope modified by partial 
covering
or by a 'warm absorber'.

On the other hand, the soft excess frequently found in EXOSAT spectra 
were  formerly
interpreted as thermal emission from the innermost regions of a viscous 
heated accretion disk (Arnaud et al. 1985).
 As suggested by Pounds et al. (1994), both 
the soft excess and the blue bump  emission may 
arise, instead, from reprocessing the hard X-rays in dense cold cloudlets 
surviving close to the 
central source. The ionizing photons absorbed in optically thick material
 will be reemitted at the black-body equilibrium temperature ($10^5-10^6$ K) 
as long as the density of the absorbing gas is sufficiently high.
Maximum temperatures of the bump component  are evaluated to about 5$\times 
10^5$ K (Walter et al. 1994).

Variability on small scale of NGC 4051 is used to reveal  the characteristics
of the emitting clouds and of the velocity field.
The observed spectral variability on scale of hours can be
explained in terms of a change in ionization parameter plus an emerging soft 
excess
(Pounds et al. 1994).
The assumption of a typical variability time scale of one hour leads to a matter 
density
of $\geq 5 \times 10^7$ \cm3 and a thickness of the photoionized gas of $\leq 
1.4 \times
 10^{14}$ cm.  Considering that the radius of the source
of the optical and UV continuum is larger than a few $10^{14}$ cm, 
 Walter et al. (1994) claim that  it can be covered in
less than 3 years if the velocity of the absorbing clouds is larger than
200 \kms.

From the spectral observations in the optical range (De Robertis \& 
Osterbrock 1984)
it  appears that there may not be a simple dichotomy between the
broad line region (BLR) and the narrow line region (NLR) in NGC 4051. 
Instead, the continuum of line widths suggests
 that the emitting regions may be
inhomogeneusly filled with clouds or filaments showing densities in a very
large range, so that the division
into a BLR and a NLR is an extreme simplification.
The asymmetry of the narrow-line profiles is consistent with radial outflow
or expansion of the gas.
The presence of strong blue wings (Veilleux 1991) favors models
 with  radial motion and a source of obscuration. 
NGC 4051 is peculiar also because  its UV spectrum is very steep and probably 
affected by intrinsic reddening. 
The presence of dust in the nucleus of NGC 4051 has been suggested by several 
groups (Walter et al 1994). 
Veilleux (1991) claims that dust is probably present and
 is the source of line asymmetry and  that the differences in profiles 
of \Hb ~and H$\alpha$ are due to reddening and/or optical depth
effects. Balmer fluxes are known to vary over periods shorter than 
one year.

The basic model of a warm absorber  (Pounds et al. 1994)  consists of
a gaseous region close to the BLR photoionized by the central radiation, 
originating absorption features in the central 
X-ray radiation. This radiation is represented by a power-law characterized 
by  a photon index 
and normalized by the intensity at 1 keV (photons $\rm cm^{-2}s^{-1} 
keV^{-1}$).
The procedure is first to fit the  power-law plus a cold absorbing column.
Then, add one or more components seeking the best fit.
The warm absorber model introduces two free parameters: the column density, N, 
and
the ionization parameter, U.
Principal absorption edges are identified with ionized carbon, nitrogen, and 
oxygen.
The effect of a warm absorber can  be interpreted also as a set of emission 
features.

 Komossa \& Fink (1997) have recently modeled the 
absorbed spectrum of NGC 4051 from ROSAT observations
in terms of warm absorption. Their excellent model can
explain most of the features observed in the  X-ray spectrum.
However, some discrepancies with a general scenario still appear, e.g., reduced 
solar
abundances are in contrast with the absence of dust, a single component warm 
absorber
is in contrast with the plurality of cloud conditions found by
 De Robertis \& Osterbrock (1984).

In this work we  consider  NGC 4051 in the light of a multi-cloud model,
 as  was found appropriate also for other galaxies (e.g. the Circinus galaxy
 model by 
Contini, Prieto \& Viegas 1998).
The clouds move radially outward the galaxy. 
A composite model which consistently accounts for the effects of the  radiation 
from the active center and of  the shocks on the emitting clouds is adopted.
Particularly, we suggest that the soft X-ray excess can be
 reproduced  by reprocessed  radiation (diffuse radiation) emitted from the
 hot slabs of gas  after being partially
absorbed by optically thick  regions throughout 
the clouds.
The consistency of the model is checked up by fitting  both the SED of the 
continuum 
in a large frequency range and the emission lines.
 
In \S 2 the general model is described. In \S 3 the soft X-ray excess 
is modeled. The results
of model calculations are compared with the observed continuum SED and
the line spectrum in \S 4 and \S 5, respectively. Final remarks
follow in \S 6.

\section{The model}

Clouds in the NLR with different physical conditions and with 
  radial outward motion are assumed by the model.  
A shock forms on the outer edge of the clouds while the power-law
radiation from the active center (a.c.) reaches the inner edge. 
The flux from the central source and the shock are the primary sources
of ionization and heating of the  gas.
The cloud  may be depicted as consisting of a large number of parallel slabs
in which the conditions within any given increment are essentially uniform.
The gas is ionized and originates a diffuse radiation field  through
recombination and lines. The intensity of the diffuse radiation depends on the 
source
function which cannot be determined unless the ionization equilibrium is already 
known (Williams 1967).
 In the source function are therefore implicit the effects of collisional 
ionization.
 The role of diffuse secondary radiation
in shocked clouds is illustrated by Viegas, Contini \& Contini (1998).
 Diffuse radiation which emerges from the heated slabs 
is  partially absorbed throughout the cloud.
We suggest that the soft X-ray excess may be produced by the reprocessed
radiation, e.g. diffuse  radiation, emitted from a group of clouds in
different physical conditions.

The SUMA code is adopted (Viegas \& Contini 1994 and references therein).
A composite ionizing spectrum with spectral index $\alpha_{UV}$ = 1.4 and
$\alpha_X$ = 0.4 is assumed for all models as for previous modeling (see 
Contini, Prieto, \& Viegas 1998).
The other input parameters, i.e., the shock velocity, \Vs, the preshock
density, \n0,  the radiation flux
intensity at 1 Ryd, \F (in number of photon $\rm cm^{-2} \, s^{-1} \, eV^{-1}$),
and  the dust-to-gas ratio by number, d/g, are chosen from the observational 
evidence and are adjusted by the modeling. 
Cosmic abundances are assumed (Allen 1973) and  the preshock magnetic field 
\B0 =  $10^{-4}$ gauss.

\section{The soft X-ray excess}

 The observational data  are
 taken from Komossa \& Fink (1997, Fig. 5). 
We assume that many clouds, at different physical conditions, contribute to the  
'warm absorption'. 
A large range of densities is considered. Actually, a continuum distribution
  of densities could  be present.
Velocities are in agreement with the observed emission line FWHM (De Robertis \&
Osterbrock 1974, Veilleux 1991). 
Some clouds are shock dominated, other are radiation dominated with radiation
intensities in the range usually found for the NLR of Seyfert  galaxies.
The input parameters of the models used to fit the X-ray excess are listed
in Table 1. 
The best fit of the X-ray excess by model calculations
is shown in Figure 1.
Models 12, 13,  and 14 are  not included in the figure
and will be discussed further on. 
 Each model represents one type of cloud with a characteristic
geometrical width, D, which is also given in Table 1. Shock dominated (SD)
 models are calculated adopting \F = 0. Notice that model 8 is a SD
model corresponding to the radiation dominated (RD) model 9. For each model,
the diffuse radiation emitted by each cloud is calculated. 
The thick solid line in Fig. 1 represent the weighted summed spectrum from the 
models. 
In the last column of Table 1 the weights,  W, adopted
for  single-cloud models in the average sum are shown. 
The weights
reflect the dilution factor $\rm (r/d)^2$ (r is the distance
from the clouds to the center and d is the distance to earth)
and the  covering factor.
It can be noticed that an acceptable fit to the observations 
is obtained by the present sample of models, considering that the
observed data are contaminated by residuals and errors. 

X-ray data at energies higher than 1.3 KeV can be well fitted by the flat
power-law radiation  from the central source, which is directly reaching the 
observer.
Lower energy photons are generally absorbed by the cloud. This can be easily
seen in the curve representing model 5, which corresponds to high density 
clouds.
The spectrum given by  model 7 shows a trend similar to the observed one,
however lacks the throats of absorption in the critical edges.
The two primary flux models have the lowest  weights. 

High density clouds are invoked to fit the deep throat at about 0.85 KeV,
which is due to the O VIII edge.
The geometrical thickness of the clouds are small, particularly for dense
clouds, indicating that fragmentation is rather strong. This is
consistent with a regime of turbulence in the presence of shocks.
The weights of the high density models 
are very high, particularly for model 3. 
In the corresponding  clouds the cooling rate is
high, due to the high density; moreover, the hot gas emitting region
is very small. Consequently, the flux emitted by each cloud is weak
and many clouds are necessary to fit the data.
This is consistent with the small D.
Model 6 with a larger D 
has also a high  weight. In this case most of the gas
inside the cloud is cold and neutral because
the intensity of the central radiation   is relatively low.

Modeling implies the choice of a composite model which is seldom unique. 
The validity of the present composite model for the warm absorber
 will  be checked in the next sections by the consistent fit to
the observed  continuum SED and to the line spectrum.

\section{The continuum}

References of the observed continuum below $10^{16}$ Hz  are given in Table 
2.
Data in the X-ray range are from Komossa \& Fink (1997).

The SED of the observed continuum is plotted in Fig. 2. 
 Optical observations integrated over the whole galaxy are not included.
For energies less than 13.6 eV, the continuum calculated by the models, 
which is essentially the sum of bremsstrahlung
radiation emitted from the gas within the clouds, roughly  fit the data.
The weights adopted to sum up the models are the same as listed in Table 1. 
Reradiation by dust in the IR depends
strongly on the shock velocity. The observed IR maximum constrains the 
dust-to-gas ratio (d/g value), while the 
frequency corresponding to the maximum  depends on the dust temperature.
The grains are heated by radiation and by collisions. Dust and gas mutual 
heating 
and cooling determine the temperature of dust which follows the temperature of 
the gas (Viegas \& Contini 1994).
For all the models  dust-to-gas ratios in the range
 1-3$\times 10^{-15}$ are adopted.

The SED of the continua corresponding to single models 1,2,3,4,6,8,9,10, and 
11
(dotted lines) and their weighted sum (solid lines) are shown in Fig.2. The
three components originating in the clouds (synchrotron emission due to Fermi 
mechanism, dust emission, and free-free emission) are shown separetely.

 It can be noticed 
that most of the models are below the lower edge of the figure because their
weight is very low. The weighted sum corresponds to the SED of model 3 the 
weight of which largely prevails.

Depending on the models, the bremsstrahlung component peaks at a different
frequency. So, the weighted average shows two peaks, one at $\sim 10^{14}$ Hz
and another at 3 $\times 10^{16}$ Hz.
  Actually, absorption by ISM peaks at 3 $\times 10^{16}$ Hz (Zombeck 1990).
In the radio range ($< 10^{10}$ Hz), the
free-free emission is higher than the observational data, which are nicely 
fitted
by synchrotron emission due to Fermi mechanism at the shock front. As happens for 
Circinus,
the bremmstrahlung emission at such low frequencies is probably absorbed.
In fact,  if
we assume an average temperature for the clouds of about 10$^4$ K, the optical
depth for free-free absorption is greater than unity for $\nu \leq 10^{11}$,
increasing at lower frequencies.

Notice, however, that the observed optical continuum is not well fitted yet.
 Moreover, reradiation by dust calculated by the models peaks at $10^{13}$ Hz, 
while the data peaks at $\sim 3 \times 10^{12}$ Hz.
Therefore, the ensemble of clouds which explain the X-ray data
is not complete, and models representing other clouds at different physical 
conditions must be included in the  multi-cloud model.
A final choice of the best fitting models will be possible after
discussing the line spectrum. In fact, modeling the line and continuum
spectra simultaneously implies cross checking of one another
until a fine tuning of the models is obtained.

\section{The optical - near-UV line spectrum}

The observed line spectrum is taken from Malkan (1986, Table 1).
A typing error crept in the published  data has been corrected
(\Hb = 31. and not 3.1, M. Malkan, private communication). 
The data are reddening corrected  adopting E(B-V) = 0.32
which represent the obscuration inside the clouds (Malkan 1986).
This is higher than the
intrinsic reddening, E(B-V)=0.08 and galactic reddening E(B-V)=0.02. 
Notice that  Walter et al. (1994) obtain E(B-V)=0.05-0.13.

The  calculated line intensities relative to H$\beta$
are compared  to the observations  in Table 3.
Radiation dominated models provide relatively high HeII/\Hb ~(e.g. model 9),
while shock dominated models provide higher [OII]/\Hb ~and 
[OIII]4363/\Hb ~(e.g. model 8).
The results presented in Table 3 indicate that both radiation-dominated
 and shock-dominated clouds should be taken into  account for the final 
multi-cloud model.
Model AV0 corresponds to the weighted average of  the single-cloud 
models accounting for the X-ray data.
This average model  gives line ratios practically  identical to model 3, 
which, in fact, largely prevails. However, the fit to the observed
line ratios is not good enough.
Therefore, models 12, 13, and 14, which are negligible in the fit of the
soft X-ray excess, are invoked to improve the fit of the line ratios
observed in the optical-near UV range.

The most noticeable features 
in the spectrum of NGC 4051 are that the ratio of the high ionization line 
widths
to the low ionization line widths is considerably smaller than for other objects
in the sample of De Robertis \& Osterbrock (1984) and that blue wings 
up to -800 \kms are present in all the forbidden line profiles (Veilleux 
1991).
Veilleux (1991) also noticed four 'shoulders'  (at -40, -110, -180,
and -350 \kms) in the observed line profiles.  The line intensity ratios 
observed by Veilleux are not included in Table 3 which
shows  the line ratios to \Hb. In fact,  the narrow and broad component of \Hb 
~could not be deblended.

The 'shoulders' indicate the velocities of emitting
clouds which  are  represented by models 14, 13, 3, in addition
to those represented by models 4 to 10. 
Model 12 represents the high velocity gas. The preshock density is high enough
to  cause the rapid cooling of the gas downstream. So,  low ionization line 
ratios relative to
\Hb ~are particularly high (Table 3), and bremsstrahlung emission 
in the X-ray domain is  low (Fig. 3).
Dust reemission is completely annihilated by the sputtering of the grains
in the immediate postshock region.
Model 13 is characterized by a low \Vs ~(100 \kms),  a low \n0 (100 \cm3), and a
low primary flux (\F = 5 $10^{10}$ units).
 These conditions generally represent the clouds either in the outer NLR or 
 in Liners  (Contini 1997). 
This model shows a too high [OIII] 4959/[OII] 3727 line ratio , 
while model 14, which is a SD model characterized by a low \Vs (50 \kms),
provides a very strong [OII]/\Hb .
The averaged spectrum (AV1) is given in the last column of Table 3 and shows an
acceptable fit  to the observations.

 The  relative contributions of models 3,12,13 and 14 to single line fluxes 
are shown in Table 4. 
As the models are distinguished particularly by the shock velocities, the results 
refer to the line profile features observed by Veilleux (1991).
Model 3 is chosen to represent the contribution of the  clouds generating the 
X-ray excess.
Interestingly, the shock velocities adopted to fit the spectra
confirm the observed prevailing FWHM  of about 200-300 \kms in the 
observed line profiles
of \Hb,  [OIII] 4363, and [OI]. Models calculated with \Vs = 100 \kms
provide 98-99 \% of the [OIII] 4959+5007 and HeII 4686 lines, 94 \% of
He II 3200, and 85-86 \% of the [NeV] 3426 and [NeIII] 3869 lines, respectively.
Models calculated  with \Vs = 50 \kms contribute essentially to
the [OII], [NII], and [SII] 6717,6730 lines. The high velocity model (12) 
contributes to all the lines, except [OIII], [NeIII], [NeV], and HeII 4686. 
Model  results are not in full agreement with
observations for all the lines, but they roughly show  how complex is the 
structure of the  NLR which
extends from the edge of the BLR to the outskirts of the galaxy.

Models 13 and 14 are  chosen also to improve the fit  of the continuum SED. 
In fact,
the mutual heating of dust and gas provides a dust temperature low enough to
settle the peak in the IR at about 3 $10^{12}$ Hz and  the trend
of model AV1 in the optical range improves the fit which was obtained by model 
AV0 (Fig. 3). Clouds  corresponding to models 12, 13, and 14 
contribute mostly to emission lines   but  not to the X-ray excess.

\section{Final Remarks}

Komossa \& Fink find that the X-ray spectrum consists of a power-law modified by 
absorption edges and an additional soft excess during the high-state in source 
flux.
Their results indicate a column density of ionized material of log N = 22.7
and a ionization parameter of log U = 0.4. The underlying power-law is in its 
steepest
observed state with photon index $\Gamma_X$ = -2.3.
 They assume that the absorber is one-component with a
gas temperature of 3 $\times 10^5$ K,  metal abundances up to 0.2 $\times$ 
solar,
electron density \ne  $\leq 3 \times 10^7$ \cm3, a thickness D $\geq 2 \times 
10^{15}$
cm, at a distance r $\geq 3 \times 10^{16}$ cm from the central power 
source , 
and no dust. Moreover, they claim that no emission line component can be fully
identified with the warm absorber. Notice that in photoionization models,
the high temperature is associated
to a low chemical abundance. However, it is well known
that the galaxies often show an abundance gradient, indicating higher abundances
in the central regions. Thus unless the depletion is due to dust, it seems 
unreal to assume such low abundances in the central region of NGC 4051.

In the previous sections we have selected the models
which consistently fit  the observed continuum in all the
frequency ranges and the line ratios.
Our results show that the so-called warm absorber is composed by many 
clouds in different physical conditions. The column density
within each cloud contributing to the soft X-ray excess
does not exceed 5$\times 10^{20}$ $\rm cm^{-2}$,
considering  a postshock compression of $\sim$ 10 for low velocity clouds
(200-300 \kms).
Comparing with the results obtained by Komossa \& Fink this
indicates that hundreds of clouds form the warm absorber.
The preshock densities span from the values which fit the NLR
to values approaching those of the BLR, i.e. between 400 and $10^7$ \cm3,
in agreement with the predictions of  De Robertis \& Osterbrock (1984) that a 
large 
range of densities characterize the emitting clouds in NGC 4051. 
Obviously, the clouds responsible of the edge absorption are the densest, 
in agreement with the density  indicated by Komossa \& Fink. 
  Moreover, they predict no dust, while  
the modeling of the continuum in the present work shows that
dust is present inside the clouds  to explain the  IR emission.
However, the dust-to-gas ratio is rather low, even lower by a factor $\geq 3$
than found for Liners by Viegas \& Contini (1994).

The central radiation flux  reaching the clouds 
ranges from  $10^{11}$ to  $10^{13}$ photons per cm$^{-2}$
s$^{-1}$ eV$^{-1}$  at the Lyman limit which are "normal"
values in the NLR of AGN (Viegas \& Contini 1994). However,
the high density clouds which explain the soft X-ray excess  at $\sim$ 1 keV
are all shock dominated. This is an interesting result which shows that the
gas is heated by the shock. Temperatures of about 6$\times 10^{5}$ K  correspond 
in fact to shocks of $\sim$200 \kms. These temperatures are in agreement 
with the temperatures
predicted by previous models (Komossa \& Fink 1997 and references therein) 
which  were explained by very strong 
radiation from the active center photoionizing and heating  a  gas
characterized by low metal abundances, in order to reach such high 
temperatures.
Our model shows that shock dominated clouds are  present in large number, 
indicating
that the central source radiation is screened, probably by the BLR clouds, 
and that the filling factor is high.
Because the high temperature  is due to shock, the fit to the 
observations is obtained with cosmic abundances.

Finally, in the present warm absorber model we assume that   some  very dense 
clouds
are characterized by relatively low velocities (Table 1, models 1, 2, and 3).
In the nuclear region of the Circinus galaxy some clouds characterized by
velocities of 250 \kms and preshock densities of 5000 \cm3 were invoked in 
order to fit the emission spectra (Contini et al. 1998).
Generally, in AGN, higher densities correspond to higher velocities
as for the clouds corresponding to model 12 (\n0 = 8000 \cm3, \Vs = 1000 \kms).
Notice that the high density low velocity clumps are characterized by a very 
small geometrical thickness in NGC 4051, therefore, they   could be identified 
with 
the debris of high density-velocity clouds from the BLR edge which 
have been fragmented by cloud collision in a turbulent regime. Fragmentation
is generally accompanied by a considerable loss of kinetic energy.

If  model 3 represents these debries, their distance from the
active center can be calculated from
$\rm  F(H\beta)_{obs}$ $\rm d^2 $ = $ \rm F(H\beta)_{calc}$ $\rm r^2$,
where $ \rm H\beta_{obs}$ ~is the absolute flux of \Hb ~observed at earth 
(Malkan 1994), 
d is the
distance of the galaxy from earth (d=14 Mpc), $ \rm H\beta_{calc}$ ~is the 
absolute
flux of \Hb ~calculated at the gaseous clump, and r is the distance of the 
clumps from the
active center. Adopting $ \rm H\beta_{obs}$ = 31.$\times 10^{-14}$ \erg 
and $ \rm H\beta_{calc}$ = 27. \erg (see Table 3), r results 1.5 pc. The high 
density
components of the warm absorber are thus located between the NLR and the BLR.

In conclusion, a multi-cloud model can explain the soft X-ray excess
in  NGC 4051, the optical emission line spectrum and the SED of the continuum.
Indeed, modeling implies the choice of some  conditions
which should actually prevail. So single-cloud models
must be considered as prototypes.
The clouds are heated and ionized by the photoionizing
flux from the active center and by the shocks. Due to the high temperature 
gas, diffuse radiation, partly absorbed throughout the clouds,
is used to explain the bump in the soft X-ray domain,
while free-free emission from lower temperature gas
 and dust reradiation from the ensemble 
of the clouds mainly fit the far IR to optical continuum.
The fit of the continuum in the IR shows that the
dust-to-gas ratio is not particularly high ($\leq 3\times 10^{-15}$).
Radio emission is well fitted by synchrotron radiation created
at the shock front by Fermi mechanism. 
Debris of high density fragmented clouds are necessary to explain
the absorption oxygen throats observed at 0.87 and 0.74 keV.
The debris are heated by shocks of about 200-300 \kms.
Low velocity ($\leq$ 100 \kms) - density (100 \cm3) clouds 
 and high velocity (1000 \kms) - density (8000 \cm3) clouds,
which are revealed from the FWHM of the line profiles,
contribute to the line and continuum spectra.

\bigskip

\noindent
{\it Acknowledgements}.
This paper was partially supported by the Brazilian funding agencies
PRONEX/Finep, CNPq, and FAPESP.

\newpage

{\bf References}

\bigskip

\vsize=26 true cm
\hsize=14 true cm
\baselineskip=18 pt
%
\def\ref {\par \noindent \parshape=6 0cm 12.5cm 
0.5cm 12.5cm 0.5cm 12.5cm 0.5cm 12.5cm 0.5cm 12.5cm 0.5cm 12.5cm}

\ref Allen, C.W. 1973 in "Astrophysical Quantities" (Athlon) 
\ref Arnaud, K.A. et al. 1985, MNRAS, 217, 105
\ref Balzano,V.A., \& Weedman,D.W. 1981, ApJ, 243,756
\ref Contini, M. 1997, A\&A, 323, 71
\ref Contini, M., Prieto, M.A., \& Viegas, S.M. 1998, ApJ, 505, 621
\ref De Robertis, M.M. \& Osterbrock, D.E. 1984, ApJ, 286, 171
\ref De Vaucouleurs,G., De Vaulcouleurs, A., Corwin, G.H.G. et al.
1991, Third Reference Catalogue of Bright Galaxies, Version 3.9
\ref De Vaucouleurs, A., Longo, G., 1988, Catalogue of Visual and Infrared
Photometry of Galaxies from 0.5 $\mu$m to 10 $\mu$m (1961-1985)
\ref Fiore, F. et al. 1992, A\&A, 262, 37
\ref Gregory,P.C. \& Condon,J.J. 1991 ApJS, 75,1011
\ref Ficarra,A., Grueff,G., \& Tomasetti,G.  1985, A\&AS, 59,255
\ref Komossa, S. \& Fink, H. 1997, A\&A, 322, 719
\ref Lebofsky,M.J. \& Rieke,G.H. 1979, ApJ, 229,111
\ref McAlary,C.W., McLaren,R.A., \& Crabtree,D.R. 1979, ApJ, 234,471
\ref Malkan, M. 1986, ApJ, 310, 679
\ref Moshir, M. et al. 1990, Infrared Astronomical Satellite Catalogs,
The Faint Source Catalog, Version 2.0
\ref Osterbrock, D.E. 1977, ApJ, 215, 733
\ref Penston,M.V., Penston,M.J., Selmes, R. A., Becklin, E. E., 
\& Neugebauer, G.  1974, MNRAS, 169,357 
\ref Pounds, K.A., Nandra, K., Fink, H.H., \& Makino, F. 1994, MNRAS, 267, 193
\ref Rieke,G.H. 1978, ApJ, 226,550
\ref Rieke,G.H. \& Low, F.J. 1972, ApJL, 176,L95
\ref Stein,W.A., \& Weedman,D.W. 1976, ApJ,205,44
\ref Soifer,B.T., Boehmer, L., Neugebauer, G., \&Sanders, D. B. 
1989, AJ, 98,766
\ref Veilleux, S. 1991, ApJ, 369, 331
\ref Viegas, S.M. \& Contini, M. 1994, ApJ, 428, 113 
\ref Viegas, S.M., Contini, M., \& Contini, T. 1998, A\&A submitted
\ref Walter, R. et al.  1994, A\&A, 285, 119
\ref Williams, R.E. 1967 ApJ, 147, 556
\ref Wisniewski,W.Z. \& Kleinmann,D.E. 1968, AJ,73,866
\ref Zombeck, M.V.  1990 in "Handbook of Space Astronomy and Astrophysics"
Cambridge University Press, p. 199

\newpage

{\bf Figure Captions}

\bigskip

Fig. 1

The X-ray spectrum of NGC 4051. Filled squares indicate the data
Single model results are indicated by numbers  which refer
to Table 1. The thick solid line represents the weighted sum
which best fits the data.

\bigskip

Fig. 2

The fit of the SED of the continuum by the ensemble of models
which form the warm absorber. Dotted lines represent
the single models and solid lines the weighted sum. 
Filled squares refer to observations in the X-ray (see Fig. 1).
Open squares indicate the observation data   at lower
frequencies.

\bigskip

Fig. 3

Same as Fig. 2 including models 12 (dotted line), model 13
(short-dashed lines), and model 14 (long-dashed lines).
The dot-dashed line shows that free-free emission is absorbed
in the radio range.
The thick solid lines refer to model AV1 and the thin solid lines
to AV0.

\newpage

\begin{table}
\centerline{Table 1}
\centerline{The input parameters of the models}
\begin{center}
\small{
\begin{tabular}{rrr rlr }\\ \hline   \\
\ model  & \n0  & \Vs &  D    &log(\F) & log(W)   \\
\  &(\cm3)& (\kms) &  (cm)& - &-   \\
\hline \\
\  1&1.(6)&300  & $<$4.5(13)&sd &-6.8  \\
\  2&5.(5)&200  & 2.(12)& sd&-4.8   \\
\  3&4.(5)&200  & 3.(13)&sd& +0.1   \\
\ 4&800&  400 & $<$2.(16)& sd&-10.4     \\  
\ 5& 800 &  420 & 5.(16)& 12 & -13.2 \\
\ 6& 800 &  300 & 2.(17)& 11 & -6.9 \\
\ 7& 600 &  300 & 1.(16) &11 & -13.0 \\
\ 8&600  &  300 & $<$1.(16)& sd &-9.4     \\
\ 9&600 &  300 & 1.(16)& 11 & -10.2  \\
\ 10&500  &400  &3.(16)&13&-10.7   \\
\ 11&400 & 200  & 7.(16)& 12& -8.4 \\
\ 12 &8000 & 1000 & 3.(17) & sd&-4.9   \\
\ 13 &100  & 100  & 3.(16) &10.3 & +4.0   \\
\ 14 &100 &  50 & $<$3(16)& sd& +3.8  \\
\hline \\
\end{tabular}}
\end{center}
\end{table}

\begin{table}
\centerline{Table 2}
\centerline{The data of the continuum}
\tiny{
\centerline{
\begin{tabular}{l ll } \\ \hline \\
 Wavelength &Frequency (Hz)  & Reference \\ \hline \\
 4400 A & 6.81 $10^{14}$ & 1 \\
 5530 A & 5.42 $10^{14}$ & 1, 2, 3 \\
 7000 A & 4.28 $10^{14}$ &  2, 3 \\
 9000 A & 3.33 $10^{14}$ & 2 \\
 1.23 $\mu$m & 2.44 $10^{14}$ & 4, 5 \\
 1.25 $\mu$m & 2.40 $10^{14}$ & 6 \\
 1.26 $\mu$m & 2.38 $10^{14}$ & 2 \\
 1.60 $\mu$m & 1.84 $10^{14}$ & 6, 7 \\
 1.66 $\mu$m & 1.81 $10^{14}$ & 4, 5 \\ 
 2.20  $\mu$m & 1.36 $10^{14}$ & 7 \\
 2.22 $\mu$m & 1.35 $10^{14}$ & 2, 3, 4, 5, 6 \\
 3.45 $\mu$m & 8.69 $10^{13}$ & 6 \\
 3.50 $\mu$m & 8.57 $10^{13}$ & 8, 9 \\
 3.54 $\mu$m & 8.47 $10^{13}$ & 3 \\
 3.65 $\mu$m & 8.21 $10^{13}$ & 4 \\
 4.65 $\mu$m & 6.45 $10^{13}$ & 4 \\
 5.00 $\mu$m & 6.00 $10^{13}$ & 9 \\
 10.5 $\mu$m & 2.86 $10^{13}$ & 9 \\
 10.6 $\mu$m & 2.83 $10^{13}$ & 6, 10 \\
 12.0 $\mu$m & 2.50 $10^{13}$ & 11, 12 \\
 21.0 $\mu$m & 1.43 $10^{13}$ & 10 \\
 25.0 $\mu$m & 1.20 $10^{13}$ & 11 \\
 60.0 $\mu$m & 5.00 $10^{12}$ & 11, 12 \\
 100.0 $\mu$m & 3.00 $10^{12}$ & 11, 12 \\
 6.18 cm&4.85 $10^9$ & 13 \\
 73.5 cm& 4.08 $ 10^8$ & 14 \\ \hline \\
\end{tabular}}}

1: De Vaucouleurs,G. et al. 1991 ; 2: Wisniewski,W.Z. \& Kleinmann,D.E. 1968;
3: De Vaucouleurs,G. et al. 1988; 4: McAlary,C.W., McLaren,R.A., \& 
Crabtree,D.R. 1979;
5: Balzano,V.A., \& Weedman,D.W. 1981; 6: Rieke,G.H. 1978;
7: Penston,M.V. et al. 1974; 8: Stein,W.A., \& Weedman,D.W. 
1976; 9: Rieke,G.H. \& Low, F.J. 1972; 10: Lebofsky,M.J. \& Rieke,G.H.
1979; 11:  Moshir, M. et al. 1990; 12: Soifer,B.T. et al. 1989; 
13: Gregory,P.C. \& Condon,J.J. 1991; 14: Ficarra,A., Grueff,G., 
\& Tomasetti,G. 1985.
\end{table}

\oddsidemargin 0.1cm
\evensidemargin 0.1cm

\vsize=24 true cm
\hsize=17 true cm

\begin{table}
\centerline{Table 3}
\centerline{The line spectra}
\tiny{
\begin{tabular}{lll llll l l l l llllll}\\ \hline   \\
\  & $\rm obs^1$  & mod & mod& mod & mod &mod &mod&mod&mod&mod& mod& mod & mod & 
mod & mod \\
\  &  &1&2&3&  4 &6&8&9 &10& 11& AV0 &12& 13 & 14 & AV1 \\
\hline \\
\ [OIII] 4959 & 0.45&1.5(-4)&0.019&0.013&0.18&0.14&3.5&7.7&0.06&3.6 
&0.013&2(-4)&2.5 & 0.075&0.76 \\
\ HeII 4686 & 0.2&1.7(-3)&2.6(-3)&1.5(-3)&0.027&3.3(-4)&0.04&0.68&0.89 
&0.85&1.5(-3)&6(-6)&0.9&0.0&0.27 \\
\ $\rm H_{\gamma}$&0.40&0.46&0.45&0.45&0.45&0.45&0.43&0.47&0.47&0.47& 
0.45&0.45&0.45&0.45&0.46  \\
\  [OIII] 4363 &$\uparrow$&3(-3)&0.15&0.08&0.06&1(-4)&0.44&0.35&0.01&0.27&0.08&  
4(-4) & 0.14 & 0.014 & 0.09 \\
\ [SII]  4072+ &  -&1.&0.09&0.06&0.47&0.05&0.48&1.2&0.0&1(-5)& 0.06 & 1.2& 2(-5) 
& 0.87 & 0.16 \\
\ [NeIII] 3869 & 0.25&0.025&0.085&0.05&0.36&7.3(-3)&1.32&4.4&0.014&0.8&0.05  & 
3(-3)& 0.74 & 0.05 & 0.26 \\ 
\ [OII] 3727+& 0.42 &4(-4)&1(-3)&7(-4)&0.6&0.046&3.7&0.04&0.0&0.023& 7(-4) & 
0.16&0.11 & 120. & 0.5 \\
\ [NeV] 3426+ &0.59&0.01&0.3&0.17&0.34&9(-5)&2.65&0.05&9.3&6.17&0.17  & 1(-6) & 
2.24 & 0.0 & 0.79 \\
\ HeII 3204 &0.1 &8(-4)&2(-3)&1.1(-3)&0.014&1.2(-4)&0.034&0.28&0.4&0.37 & 
1.1(-3) & 0.06 & 0.37 & 0.0 & 0.12 \\
\ $\rm H\beta^2$ & - &1.6(3)&18.&27.&0.063&81.7&5(-3)&0.45&0.62&0.67 &-&4(5)  
&1.2(-3) & 2.5(-5) & - \\
\hline \\
\end{tabular}}

$^1$  corrected ($\rm E_{(B-V)}$=0.32)

$^2$ in  \erg
\end{table}

\begin{table}
\centerline{Table 4}
\centerline{The \% of the different components in single line fluxes}
\begin{center}
\begin{tabular}{lll ll}\\ \hline   \\
\  & mod & mod & mod& mod \\
\  &  3&12 & 13 & 14  \\ \hline \\
\ [SII] 6730   & 7.2  & 6.7  & 2.0 & 84.  \\
\ [SII] 6717   & 2.64   & 2.4  & 1.7 & 93. \\
\ [NII] 6583   & 2.55  & 37.4 & 12.8 & 47.2 \\
\ [OI] 6300+& 92.2 & 6.4 & 0.0 & 1.37 \\
\ [OIII] 4959 & 1.15 & 2.6(-3) & 98.8 & 0.04  \\
\ HeII 4686 & 0.37  & 7.4(-3) & 99.6 & 0.0 \\
\ [OIII] 4363 &56. & 0.04 & 43.7  & 0.057  \\
\ [SII]  4072+ & 24.7  & 73.2  & 3.6(-3)  & 2.1  \\
\ [NeIII] 3869 & 13.2  & 0.12  & 86.6  &  0.078  \\
\ [OII] 3727+& 0.089  & 3.0  & 6.0  & 90.7  \\
\ [NeV] 3426+ &14.6  & 0.0  & 85.4  & 0.0  \\
\ HeII 3204 &0.62 & 5.0  & 94.2  & 0.0  \\
\ $\rm H\beta$ & 62.6  & 9.27  & 27.8  & 0.37  \\ 
\hline \\
\end{tabular}
\end{center}
\end{table}

\newpage

\topmargin 0.01cm
\oddsidemargin 0.01cm
\evensidemargin 0.01cm

\begin{figure}
\begin{center}
\centerline{Fig. 1}
\mbox{\psfig{file=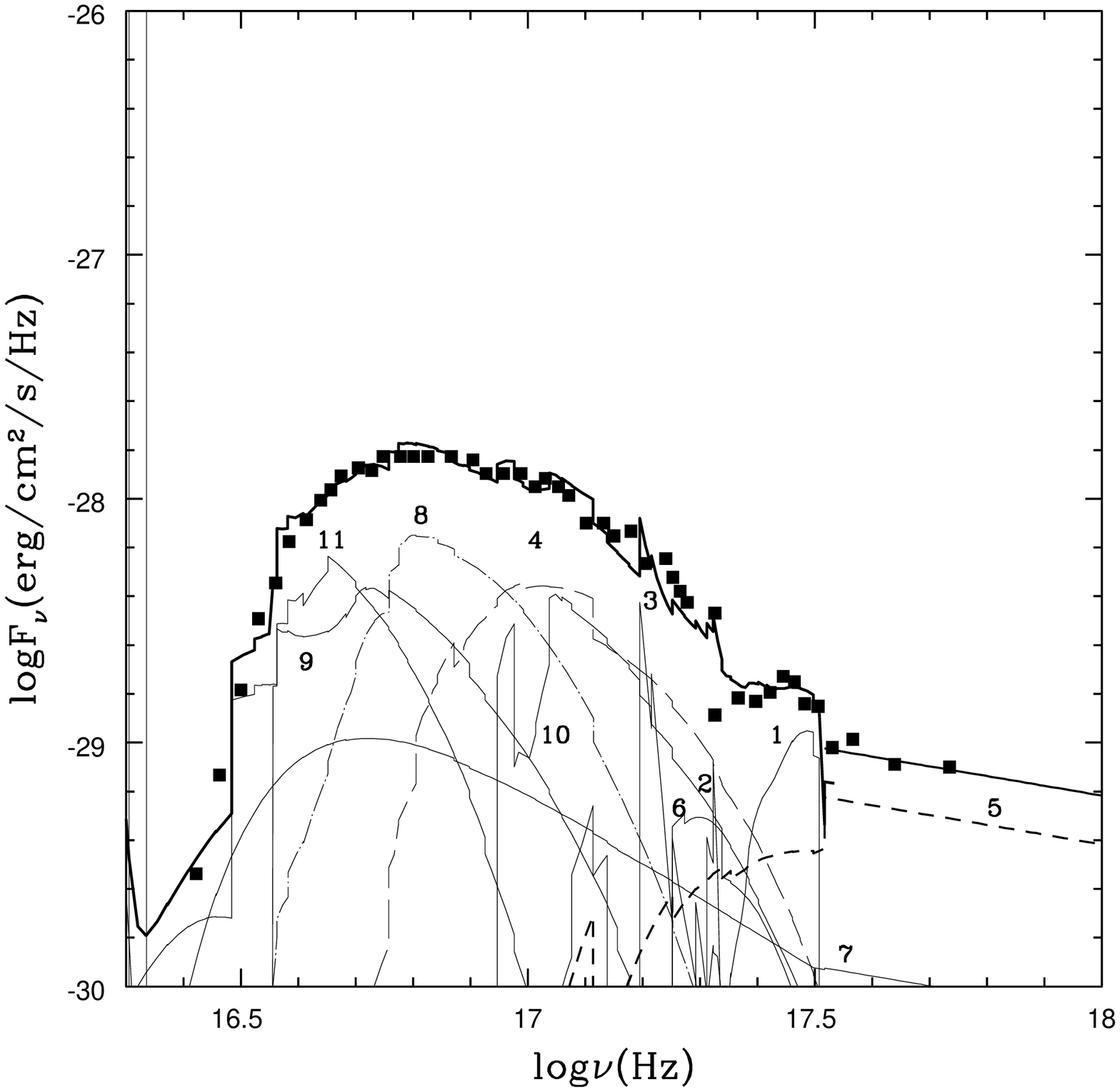,clip=,height=12.cm,width=12.cm}}
\end{center}
\end{figure}

\begin{figure}
\begin{center}
\centerline{Fig. 2}
\mbox{\psfig{file=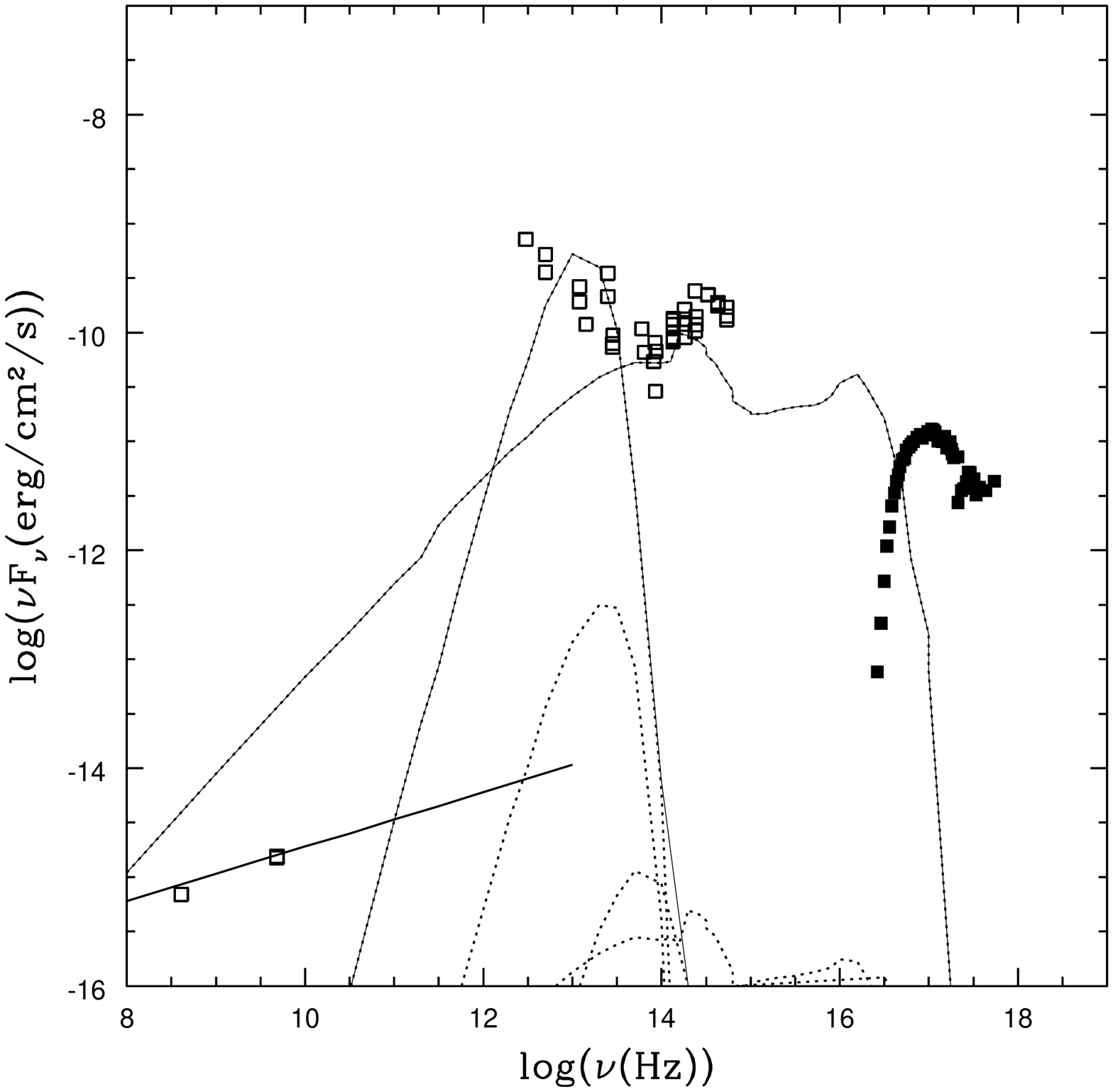,clip=,height=14.6cm,width=14.6cm}}
\end{center}
\end{figure}

\begin{figure}
\begin{center}
\centerline{Fig. 3}
\mbox{\psfig{file=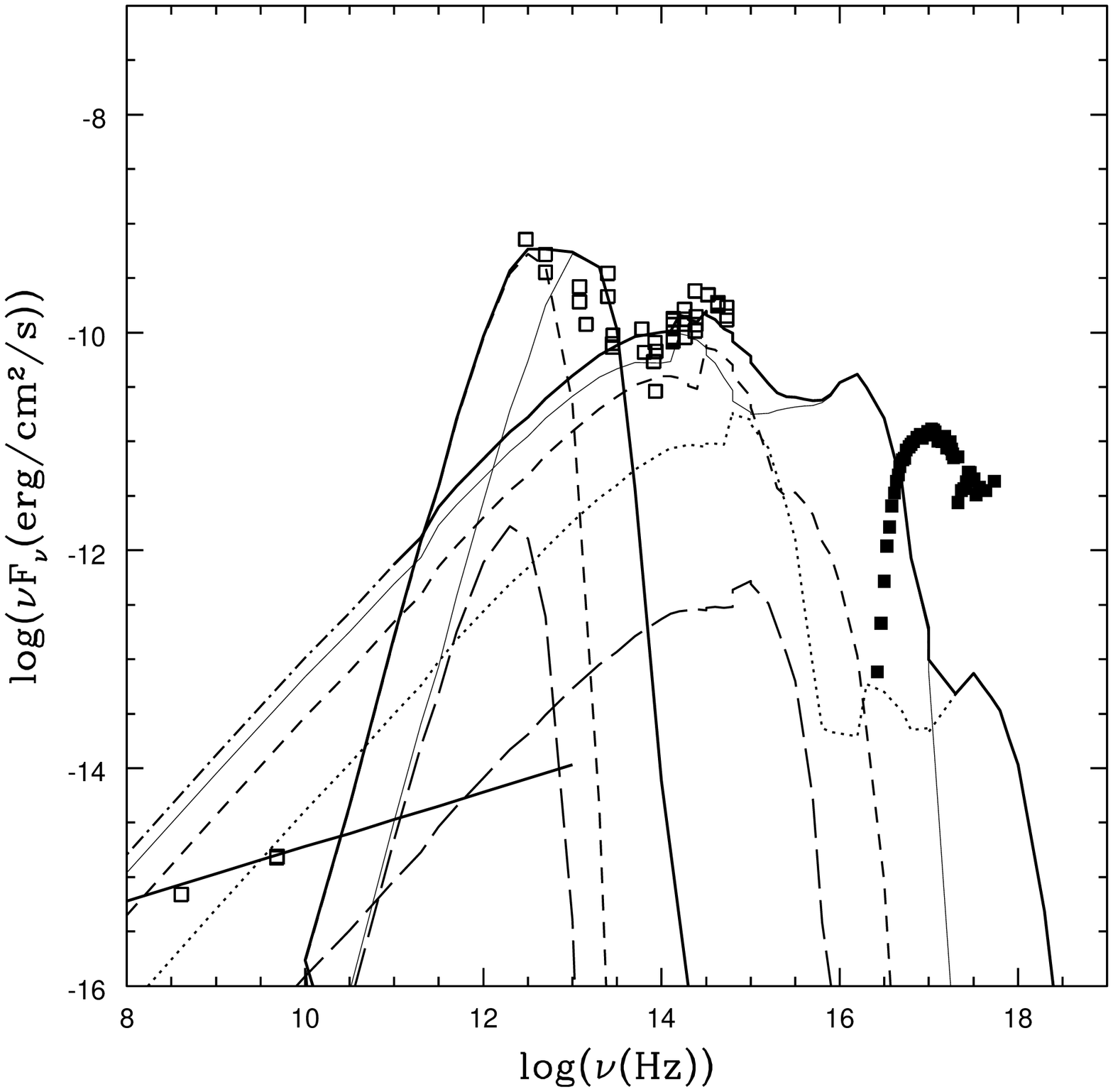,clip=,height=14.6cm,width=14.6cm}}
\end{center}
\end{figure}

\end{document}